# Cryptography, Quantum Computation and Trapped Ions

Richard J. Hughes
Los Alamos National Laboratory
Los Alamos, NM 87545, USA

ABSTRACT

The significance of quantum computation for cryptography is discussed. Following a brief survey of the requirements for quantum computational hardware, an overview of the ion trap quantum computation project at Los Alamos is presented. The physical limitations to quantum computation with trapped ions are analyzed and an assessment of the computational potential of the technology is made.

## 1. Introduction

Over the past decade information theory has been generalized to include quantum mechanical systems. (A two-level quantum system has come to be known as a qubit in this context.) The additional freedom introduced with the quantum mechanical superposition principle has opened up a variety of capabilities that go well beyond those of conventional information techniques. For example, quantum cryptography [1, 2] allows two parties to generate a secret key even in the presence of eavesdropping. But perhaps the most remarkable capabilities have been predicted in the field of quantum computation,[3, 4] which has seen tremendous growth since 1994. There are three reasons for this increased interest: the invention of quantum algorithms with compelling applications to cryptography; realistic hardware proposals and related experiments; and the development of quantum error correcting codes. In this paper I shall describe why quantum computation is now such an active field; why it is experimentally difficult; and the prospects for quantum computation with trapped ions.

Feynman [3] (see also Benioff [5]) investigated whether quantum mechanics imposes any intrinsic limitations on conventional computation, and he demonstrated that reversible Boolean logic could be implemented through quantum mechanical unitary transformations. However, it was Deutsch [6] who first suggested that by also using non-Boolean unitary operations the quantum superposition principle could be exploited to achieve greater computational power than with conventional computation. But it was not until the work of Shor [7] in 1994 that this "quantum parallelism" was shown to offer an efficient solution of an interesting computational problem. Building on earlier work of Simon [8], Shor invented polynomial-time quantum algorithms for solving the integer factorization and discrete logarithm problems.[7] The difficulty of solving these two problems with conventional computers underlies the security of much of modern public key cryptography.[9] Shor's algorithms are sufficiently compelling that the daunting scientific and technological challenges involved in practical quantum computation are now worthy of serious experimental study. (Since 1994 other interesting quantum algorithms, such as Grover's "database search" algorithm,[10] have been invented, but in this paper I shall concentrate on the implications of quantum factoring.)

Although experimental quantum computation is in its infancy, there is a very promising hardware concept using the quantum states of laser-cooled ions in an electromagnetic trap.[11] Cirac and Zoller showed that such systems have the necessary characteristics to perform quantum computation. The relevant coherence times can be adequately long; mechanisms for performing the quantum logic gate operations exist; and a high-probability readout method is possible. (For a detailed description see Reference 12.) Several groups, including our own, [13] are now investigating quantum computation with trapped ions. A single logic operation using a trapped beryllium ion has been demonstrated.[14] However, even algorithmically simple computations will require the creation and controlled evolution of entangled quantum states that are far more complex than have so far been achieved experimentally. It is therefore important to quantify the extent to which trapped ions could allow the quantum engineering of the complex states required for quantum computation.[15] Furthermore, by characterizing the way in which the precision of quantum operations depends on experimental parameters, it will be possible to determine how quantum error correction schemes and fault-tolerant methods can best be applied. These concepts hold out the prospect of unlimited quantum computation with imperfect physical implementations, if certain precision thresholds can be attained.[16]

The rest of this paper is organized as follows. In Section 2 we review the cryptographic implications of quantum factoring. In Section 3 we describe the Cirac-Zoller scheme for ion trap quantum computation, and Section 4 is devoted to a description of the different qubit schemes possible with trapped ions. Sections 5 and 6 contain estimates of the limits to quantum computation with the two classes of qubits. In Section 7 we compare the bounds obtained in Sections 4 and 5 with the requirements of quantum factoring. Finally in Section 8 we present some conclusions.

## 2. Factoring, Public Key Cryptography and Quantum Computation

Every integer can be uniquely decomposed into a product of prime numbers. Most integers are easy to factor because they are products of small primes, but large integers (hundreds of digits in length) that are products of two, distinct, comparably-sized primes can be very difficult to factor with conventional computers.[17] For example, in 1994 the 129-digit number known as RSA129 [18] required 5,000 MIPS-years of computer time over an 8-month period, using more than 1,000 workstations, to determine its 64-digit and 65-digit prime factors.[19] (By convention one MIPS-year is about $3 \times 10^{13}$ instructions. Current workstations are rated at 200–800 MIPS.) The perceived difficulty of factoring with conventional computers underlies the security of widely-used public key cryptosystems. It is often necessary to ensure that encrypted information remains secure for decades, but when encrypted information is transmitted we must assume that it can be monitored and saved for future analysis by eavesdroppers. A quantum computer (QC) using Shor's algorithm at a clock speed of 100 MHz would have factored RSA129 in only a few seconds. So the possibility that quantum computers could become feasible is not just a potential challenge to the use of public key cryptography in the future, but is a concern for the use of these cryptosystems today.

The RSA cryptosystem, [20] is based on the following computationally difficult problem:

**RSA problem**: Given an integer $N$ that is a product of two distinct primes, $p$ and $q$, an integer $e$ such that $g.c.d.(e, (p – 1)(q – 1)) = 1$, and an integer $C$, find the integer $M$ such that, $C = M^e \bmod N$.

(Here "$g.c.d$" denotes "greatest common divisor," and "mod $N$" indicates that arithmetic is being performed modulo $N$. Solving this problem is conjectured to be equivalent to factoring.)

To understand the significance of the RSA problem we first introduce Euler's totient function, which for an integer $m$ is defined as,

$$\phi(m) = \text{number of integers less than } m \text{ that are relatively prime to } m. \qquad (1)$$

Thus, for a prime, $p$, $\phi(p) = (p – 1)$, and for composite moduli of the form, $N = p.q$, introduced above we have,

$$\phi(N) = (p – 1)(q – 1) \; . \qquad (2)$$

We will also need the following theorem of Euler, which states that for any integer $x$, relatively prime to $m$, i.e. $g.c.d.(x, m) = 1$,

$$x^{\phi(m)} = 1 \bmod m \; . \qquad (3)$$

Therefore, we can solve the RSA problem if we can find the integer $d$ defined by

$$d = e^{\phi(\phi(N)) - 1} \bmod \phi(N) \; , \qquad (4)$$

because then

$$C^d = M^{ed} = M^{k\phi(N) + 1} = M \bmod N \; , \qquad (5)$$

by Euler's theorem, Eq.(3), for some integer $k$.[21] Clearly, if we know $\phi(N)$ we can find the integer $d$, and we can determine $\phi(N)$ if we know the factors, $p$ and $q$, of $N$. Thus the RSA problem can be solved if we can factor the modulus $N$, which as we will see is computationally hard.

In the RSA cryptosystem the above problem is used to provide cryptographic security as follows. Alice wishes to send a (plaintext) message $M$ (a large integer) to Bob, but wants to be sure that the eavesdropper Eve cannot read the message. So:
a) Bob generates two large, distinct (secret) primes, $p$ and $q$;
b) He computes their product, $N$, and the integer $\phi(N)$;
c) Bob selects an integer $e$, such that $g.c.d.(e, \phi(N)) = 1$, and computes the integer $d$ as above;
d) Bob publishes his public key, comprised of the modulus $N$ and encrypting exponent, $e$, but keeps his private key (decrypting exponent) $d$ (as well as $p$ and $q$) secret;
e) Alice uses Bob's public key to compute the ciphertext $C = M^e \bmod N$, and sends $C$ to Bob;
f) Bob recovers Alice's message, $M$, using his secret decrypting exponent, $d$, as above.

We must assume that Eve passively monitors Alice's transmission, but if Eve wants to decrypt Alice's communication she is faced with the computationally hard problem of factoring Bob's modulus, $N$. Alice and Bob may wish to be sure that their communication will remain secret for many years, even decades, so they must assess how much computational power Eve could apply to factoring the modulus $N$, not just today but for many years into the future.

One way to factor $N$ would be to perform trial divisions by all primes less than $N^{1/2}$. However, this would require $O(\exp[l])$ divisions, where $l$ is the number of bits in the binary representation of $N$, and so the amount of computational work required would grow exponentially with the size of the modulus. Modern factoring algorithms (including Shor's quantum factoring algorithm) use a different strategy:[22] they search for non-trivial solutions, $y$, of Legendre's conguence,

$$y^2 = 1 \bmod N \quad , \tag{6}$$

from which we have,

$$(y + 1)(y - 1) = 0 \bmod N \quad . \tag{7}$$

Then the factors of $N$ will be distributed between the two parentheses, and can be found using

$$g.c.d.(y \pm 1, N) = \text{factor of } N \quad . \tag{8}$$

The General Number Field Sieve (GNFS) algorithm [23] is the best algorithm is use today for factoring large integers. It is much more efficient than trial division, with a sub-exponential run-time growth, $O[1.923 l^{1/3}(\ln(l))^{2/3}]$, and is well-adapted to distributed processing. This algorithm was recently used to factor the 130-digit number known as RSA130 [24] in 500 MIPS-years of computer time. From this result we can estimate the amount of computer time that would be required to factor (using the GNFS) the much larger moduli that are used in RSA cryptosystems today.

| Size of modulus (bits) | 512 | 1,024 | 2,048 | 4,096 |
|---|---|---|---|---|
| Factoring time (MIPS-years) | $2 \times 10^4$ | $2 \times 10^{12}$ | $6 \times 10^{22}$ | $3 \times 10^{36}$ |

Table 1: Factoring times in MIPS-years for various moduli using the GNFS.

If we assume that Eve can obtain the dedicated use of 1,000 workstations we can convert the results of Table 1 into factoring times for various future years. We assume: that each workstation in 1997 is rated at 200 MIPS; that there are no algorithmic developments beyond the GNFS; and that in the future Eve uses state of the art systems with processing powers given by "Moore's Law." (The observation that the power of successive processor generations increases by a factor of two approximately every 18 months is known as "Moore's law".)

| Size of modulus (bits) | 1,024 | 2,048 | 4,096 |
|---|---|---|---|
| Factoring time in 1997 | $10^7$ years | $3 \times 10^{17}$ years | $2 \times 10^{31}$ years |
| Factoring time in 2006 | $10^5$ years | $5 \times 10^{15}$ years | $3 \times 10^{29}$ years |
| Factoring time in 2015 | 2,500 years | $7 \times 10^{13}$ years | $4 \times 10^{27}$ years |
| Factoring time in 2024 | 38 years | $10^{12}$ years | $7 \times 10^{25}$ years |
| Factoring time in 2033 | 7 months | $2 \times 10^{10}$ years | $10^{24}$ years |
| Factoring time in 2042 | 3 days | $3 \times 10^8$ years | $2 \times 10^{22}$ years |

Table 2: Projected future factoring times using the GNFS for various moduli using 1,000 workstations.

We see from Table 2 that 1,024-bit moduli are probably inadequate for information that must remain secure for three decades or more. However, because of the exponential growth of the factoring time, moduli that have only twice as many bits provide ample protection for data that must be safe for more than four decades, in spite of the assumed exponential growth of processor power. But if Eve should develop a quantum computer even larger moduli would be inadequate.

To factor an $l$-bit integer, $N$, Shor's quantum factoring algorithm requires a classical integer, $x$, that is relatively prime to $N$, and the computation of the period of the function [25]

$$f(a) = x^a \bmod N, \quad a = 0, 1, \ldots N^2-1 \quad . \tag{9}$$

From the period of this function the order, $r$, of $x$ can be determined. The order is the smallest integer, $r$, for which

$$x^r = 1 \bmod N \quad . \tag{10}$$

If this order is even, the congruence

$$(x^{r/2}-1)(x^{r/2}+1) = 0 \bmod N \quad , \tag{11}$$

can be used to factor $N$ as described above.

Shor's algorithm therefore requires one $2l$-bit register to hold the argument of the function, $f$; an $l$-bit register to hold the function values, and some additional memory to allow reversible computation of the function. (The computation of the quantum Fourier transform to determine the order, $r$, involves an insignificant number of quantum gate operations in comparison with the computation of the function $f$.)

The amount of additional memory and number of quantum logic gates is somewhat dependent on the specific implementation of the algorithm,[26] but in our recent improved version[27] Eve would need a QC with $L$ qubits of memory and $n_g$ quantum logic operations, with

$$L = 5l + 4$$
$$n_g = 25l^3 + O(l^2) \qquad (12)$$

to factor an *l*-bit modulus. In contrast to the (sub)exponential run-time growth of the classical GNFS factoring algorithm, the quantum algorithm has a dramatically slower, polynomial, $O(l^3)$, growth. (The $l^3$-dependence can be understood as arising from the (conditional) multiplication of 2*l* classical *l*-bit integers to build the function, *f*. Each of the multiplications requires $O(l^2)$ bit-additions (using "elementary school" multiplication) that can be reduced to elementary quantum logic operations.) If we assume a nominal clock speed of 100 MHz for Eve's QC we obtain the following quantum factoring times:

| Size of modulus (bits) | 512 | 1,024 | 2,048 | 4,096 |
|---|---|---|---|---|
| Quantum memory (qubits) | 2,564 | 5,124 | 10,244 | 20,484 |
| Number of quantum gates | $3 \times 10^9$ | $3 \times 10^{10}$ | $2 \times 10^{11}$ | $2 \times 10^{12}$ |
| Quantum factoring time | 33 seconds | 4.5 minutes | 36 minutes | 4.8 hours |

Table 3: Quantum factoring times of various moduli on a hypothetical 100-MHz QC.

The quantum factoring algorithm used for these estimates is not optimized and significant improvements are possible. Nevertheless, we see from Table 3 that even 4,096-bit moduli would provide inadequate security against this hypothetical QC. Moreover, because of the slow polynomial growth of the quantum factoring algorithm's run-time, Alice and Bob cannot easily ensure security against possible future quantum attacks by making modest increases in the size of the modulus: they would need exponentially larger moduli. (For a 20-year quantum factoring time the modulus would need to be ~ 100,000 bits long, using our algorithm and Eve's hypothetical QC. Such large moduli would make encryption and decryption prohibitively slow.)

From this simple analysis we see that information encrypted with moduli offering ample security for many decades against conventional factoring attacks would be rendered retroactively vulnerable to a future quantum attack. However, because of the very large numbers of qubits and the long coherence times required we cannot yet state that quantum computation factoring of cryptographically significant integers will ever be possible. But the possibility that quantum factoring might become feasible in 20 years time (say) should be a serious concern for public-key cryptography today. Furthermore, we cannot dismiss quantum computation as a potential factoring threat that will be superceded by conventional computation with time. Unless a polynomial-time conventional factoring algorithm is invented, the future exponential growth of conventional processor power is easily offset by small increases in the size of public key moduli.

Finally we note that the potential power advantage of quantum computation over conventional does not come from the intrinsic speed of a QC's logic operations, but is instead derived from the parallelism afforded by the superposition principle. Indeed, a QC would have no speed advantages over conventional computers for most "number crunching" problems, and so we are unlikely to see a "general purpose" quantum computer. At present there are only a few types of problem known for which a QC would be advantageous. Nevertheless, if the only problem that could be solved on a QC was integer factorization, this would be reason enough to build a one.

### 3. Quantum Computational Hardware

Quantum algorithms such as quantum factoring can be built up from sequences of elementary quantum logic operations ("gates"). Because the Schrodinger equation has time-reversal symmetry, quantum logic must also be reversible. i.e. the input of a logic operation must be recoverable from the output. It is known that all conventional Boolean operations can be constructed from three reversible logic operations: *NOT*, controlled-*NOT* (*CNOT*), and controlled-controlled-*NOT* (*CCNOT*).[3] When implemented on qubits, with each qubit's Hilbert space having a basis $\{|0\rangle, |1\rangle\}$, these operations have the form:

$$NOT_i : |a\rangle_i \to |\bar{a}\rangle_i$$
$$CNOT_{ij} : |a\rangle_i |b\rangle_j \to |a\rangle_i |a \oplus b\rangle_j \quad (13)$$
$$CCNOT_{ijk} : |a\rangle_i |b\rangle_j |c\rangle_k \to |a\rangle_i |b\rangle_j |(a \wedge b) \oplus c\rangle_k$$

where the subscripts *i, j, k* denote the qubits being acted upon; "⊕" denotes the logical XOR operation or addition mod 2; "∧" is the logical AND operation; and *a, b, c* = 0 or 1. However, to realize the advantages of quantum parallelism an operation that creates superpositions is required, such as

$$V_i(\theta, \phi) : \begin{array}{l} |0\rangle_i \to \cos(\theta/2)|0\rangle_i - i\exp(i\phi)\sin(\theta/2)|1\rangle_i \\ |0\rangle_i \to \cos(\theta/2)|1\rangle_i - i\exp(-i\phi)\sin(\theta/2)|0\rangle_i \end{array} \quad . \quad (14)$$

Using this operation we can readily see the complexities involved in quantum computation, because acting on the two-qubit state |0>|0> with the two gate sequence, *CNOT.V*, produces an entangled (non-classical) state:

$$CNOT_{ij} \cdot V_i(\pi/2, \pi/2)|0\rangle_i |0\rangle_j = 2^{-1/2}\left(|0\rangle_i |0\rangle_j + |1\rangle_i |1\rangle_j\right) \quad . \quad (15)$$

From the foregoing we conclude that there are three essential requirements for quantum computation hardware. Firstly, it must be possible to prepare multiple qubits, adequately isolated from interactions with their environment for the duration of computation, in an addressable form. Secondly, there must be an external drive mechanism for performing the requisite quantum logic operations, which requires the careful and precise control of the qubits' phases. And thirdly, there must be a readout mechanism for measuring the state of each qubit at the end of the computation. It is clear that it is much easier to write down a sequence of quantum logic operations than it is to perform them in the laboratory. Nevertheless, the above conditions can be satisfied with trapped ions.

(Other quantum computational technologies that are being studied include nuclear magnetic resonance[28] and cavity quantum electrodynamics.[29])

In an ion trap quantum computer a qubit would comprise two long-lived internal states, which we shall denote $|0\rangle$ and $|1\rangle$, of an ion isolated from the environment by the electromagnetic fields of a linear radio-frequency quadrupole (RFQ) ion trap. Many different ion species are suitable for quantum computation, and several different qubit schemes are possible, as we shall see below. For example, at Los Alamos we are developing an ion-trap quantum-computer experiment using calcium ions, with the ultimate objective of performing multiple gate operations on a register of several qubits (and possibly small computations) in order to determine the potential and physical limitations of this technology.[13] We have chosen calcium ions for the convenience of the wavelengths required. The heart of our experiment is a linear radio-frequency quadrupole (RFQ) ion trap with cylindrical geometry in which strong radial confinement is provided by radio-frequency potentials applied to four "rod" electrodes and axial confinement is produced by a harmonic electrostatic potential applied by two "end caps."

After Doppler cooling on their 397-nm *S-P* transition, several calcium ions will become localized along the ion trap's axis because their recoil energy (from photon emission) is less than the spacing of the ions' quantum vibrational energy levels in the axial confining potential. Although localized to distances much smaller than the wavelength of the cooling radiation, the ions nevertheless undergo small amplitude oscillations. Their lowest frequency mode is the axial center of mass (CM) motion in which all the ions oscillate in phase along the trap axis. The frequency of this mode, whose quantum states will provide a computational "bus," is set by the axial potential. The inter-ion spacing is determined by the equilibrium between this axial potential, which tends to push the ions together, and the ions' mutual Coulomb repulsion. For example, with a 200-kHz axial CM frequency, the inter-ion spacing is on the order of 30 µm. After this first stage of cooling, the ions form a "quantum register" in which one qubit can be addressed (with a suitable laser beam) in isolation from its neighbors. We have determined that more than 20 ions can be held in an optically addressable configuration. However, before quantum computation can take place, the quantum state of the ions' CM mode must be prepared in its quantum ground state.

Because of the long radiative lifetime of the metastable 3*D*-states (~1 s), the *S-D* electric quadrupole transition in calcium ions has such a narrow width that it displays upper and lower sidebands separated from the central frequency by the CM frequency. With a laser that has a suitably narrow linewidth, tuned to the lower sideband, an additional stage of laser cooling (beyond Doppler cooling) can be used to prepare the "bus" qubit (CM vibrational mode) in its lowest quantum state ("sideband cooling"). On completion of this stage, the QC would have all qubits in the $|0\rangle$ state, ready for quantum computation. (This second stage of cooling could also be performed with Raman transitions.)

The quantum state of the register of ions will then be manipulated by performing quantum logical gate operations that will be effected by directing a laser beam at individual ions for prescribed times. The laser-ion interaction will coherently change the state of the qubit through the phenomenon of Rabi oscillations. (Several different types of transition are possible.) As we will see below, the *CNOT* operation can be effected with the help of the quantum states of the ions' CM motion to convey quantum information from one ion to the other.

On completion of the quantum logic operations the result of the quantum computation can be read out by turning on a laser that drives the transition between the |0> state and another ionic level that decays rapidly back to |0>. An ion in the |0> state will then fluoresce, whereas an ion in the |1> state will remain dark. So, by observing which ions fluoresce and which are dark, a bit value can be obtained. We have recently succeeded in trapping calcium ions in our ion trap and imaging them with a charge-coupled device (CCD) camera. This is the first step toward creation of a quantum register.[13]

### 4. Trapped ion qubits

In addition to the two states, |0> and |1> comprising each ionic qubit, in an ion trap QC there is also a computational "bus" qubit formed by the ground, |g>, and first excited state, |e>, of the ions' CM axial vibrational motion, which is used to perform logic operations between qubits. By virtue of energy conservation (and possibly other selection rules) it is possible to perform two types of coherent operations on a qubit, using laser pulses directed at an ion: on-resonance transitions that change only an ion's internal state ("$V$" pulses); and red-sideband transitions (detuned from resonance by the CM frequency) that change both the qubit's internal state and the CM quantum state ("$U$" pulses). The $V$-pulse Hamiltonian for a particular ion is,

$$H_V = \frac{\hbar\Omega}{2}\left[e^{-i\phi}|1\rangle\langle 0| + e^{i\phi}|0\rangle\langle 1|\right] \quad , \tag{16}$$

and the $U$-pulse Hamiltonian is,

$$H_U = \frac{\hbar\eta\Omega}{2\sqrt{L}}\left[e^{-i\phi}|1\rangle\langle 0|a + e^{i\phi}|0\rangle\langle 1|a^\dagger\right] \quad . \tag{17}$$

Here $\Omega$ is the Rabi frequency, $\phi$ is the phase of laser drive, $\eta$ is the Lamb-Dicke parameter (characterizing the strength of the interaction between the laser and the ions' oscillations), $L$ is the number of ions, and $a$ ($a^\dagger$) is the destruction (creation) operator for quanta of the CM motion, satisfying

$$a|g\rangle = 0 \quad , \quad a^\dagger|g\rangle = |e\rangle \quad , \quad [a,a^\dagger] = 1 \quad . \tag{18}$$

The unitary operations effected by applying these Hamiltonians to the $m$-th qubit for a duration given by a parameter $\theta$ and phase $\phi$ are:

$$V_m(\theta,\phi): \begin{matrix} |0\rangle_m \to \cos(\theta/2)|0\rangle_m - ie^{i\phi}\sin(\theta/2)|1\rangle_m \\ |1\rangle_m \to \cos(\theta/2)|1\rangle_m - ie^{-i\phi}\sin(\theta/2)|0\rangle_m \end{matrix} \quad , \tag{19}$$

and

$$U_m(\theta,\phi): \begin{matrix} |0\rangle_m|e\rangle \to \cos(\theta/2)|0\rangle_m|e\rangle - ie^{i\phi}\sin(\theta/2)|1\rangle_m|g\rangle \\ |1\rangle_m|g\rangle \to \cos(\theta/2)|1\rangle_m|g\rangle - ie^{-i\phi}\sin(\theta/2)|0\rangle_m|e\rangle \end{matrix} \quad . \tag{20}$$

To perform logic operations on the qubits an additional red-detuned operation involving the transition from |0> to an auxiliary level, |aux>, in each qubit is required, with Hamiltonian

$$H_U^{aux} = \frac{\hbar\eta\Omega}{2\sqrt{L}}\left[e^{-i\phi}|aux\rangle\langle 0|a + e^{i\phi}|0\rangle\langle aux|a^\dagger\right] \quad , \tag{21}$$

with associated unitary operation $U_m^{aux}(\theta,\phi)$. For example, the controlled-sign-flip (*CSF*) operation between two qubits, *c* and *t*

$$CSF_{ct}: \begin{array}{l} |0\rangle_c|0\rangle_t \to |0\rangle_c|0\rangle_t \\ |0\rangle_c|1\rangle_t \to |0\rangle_c|1\rangle_t \\ |1\rangle_c|0\rangle_t \to |1\rangle_c|0\rangle_t \\ |1\rangle_c|1\rangle_t \to -|1\rangle_c|1\rangle_t \end{array} \quad , \tag{22}$$

can be accomplished with the sequence of three *U*-pulses of appropriate duration:

$$CSF_{ct} = U_c(\pi,0)U_t^{aux}(2\pi,0)U_c(\pi,0) \quad . \tag{23}$$

From this operation a *CNOT* gate can be produced as

$$CNOT_{ct} = V_t(\pi/2,\pi/2)CSF_{ct}V_t(\pi/2,\pi/2) \quad . \tag{24}$$

The speed of *U*- and *V*-pulse transitions is determined by the Rabi frequency, $\Omega$, which is proportional to the square root of the laser intensity. But the *U*-pulses are slower than the *V*-pulses because they must put the ions' center of mass into motion, which is a slower process with more ions, and moreover the Lamb-Dicke parameter, $\eta$, is less than one. Because of their slowness (smallness of the coupling) the *U*-operations are the rate-limiting quantities to quantum logic operations. It is therefore desirable to drive these transitions as quickly as possible. However, the laser intensity cannot be made arbitrarily large, in order to avoid driving a *V*-transition, for instance. In the following we shall only count the duration of the *U*-pulses to the computational time.

There are two classes of candidates for the qubit levels. The first category occurs in ions such as $Hg^+$, $Sr^+$, $Ca^+$, $Ba^+$ and $Yb^+$ with first excited states that are metastable, with lifetimes ranging from 0.1 s ($Hg^+$), 0.4s ($Sr^+$), 1s ($Ca^+$); 1 min ($Ba^+$) and even 10 years ($Yb^+$). A qubit is comprised of an ion's electronic ground (S) state (|0>), and a sublevel (|1>) of the metastable excited state (a *D*-state in Hg, Ca or Ba; an *F*-state in Yb). The advantage of this scheme is that it requires only a single laser beam to drive the qubit transitions, which greatly simplifies the optics of ion addressing. However, the disadvantage of this scheme is that it requires optical frequency stability of the laser drive that effects coherent transitions between the qubit levels.

Alternative qubit schemes use hyperfine sublevels of an ion's ground state, or even Zeeman sublevels in a small magnetic field for ions with zero nuclear spin, with transitions between the qubit levels driven by Raman transitions. The advantages of this type of scheme are that the qubit states can be much longer-lived than the metastable state qubits; only radio frequency stability is required (corresponding to the frequency difference between the sublevels); and there are many more possible choices of ion ($Be^+$, $Ca^+$, $Ba^+$ and $Mg^+$ for example). Disadvantages are that addressing of the qubits is more complex owing to the requirement for two laser beams; and the readout is more involved than with metastable state qubits.

During quantum computation it is essential that a QC evolves through a sequence of pure quantum states, prescribed by some quantum algorithm. In general there will be some time scale required for a particular computation, and other time scales characterizing the processes that lead to the loss of quantum coherence. By estimating these time scales we can determine if ion trap QCs have the necessary preconditions to allow quantum computation to be performed, and which systems are most favorable. Furthermore, certain decoherence mechanisms become more pronounced with larger numbers of qubits, and there are technological limits to the number of qubits that can be held and addressed. Therefore, there are also memory (space) limitations to quantum computation, as well as time limitations, and it will be important to determine how to optimize quantum algorithms to make best use of the available resources. In our experiment we have determined that more than 20 ions can be held in a linear configuration and optically addressed with minimal cross-talk, using available technology.[13]

The various decoherence mechanisms can be separated into two classes: fundamental or technical. The former are limitations imposed by laws of Nature, such as the spontaneous emission of a photon from a qubit level, or the breakdown of the two-level approximation if a qubit transition is driven with excessive laser power. The technical limits are those imposed by existing experimental techniques, such as the "heating" of the ions' CM vibrational mode, or the phase stability of the laser driving the qubit transitions. One might expect that these limitations would become less restrictive as technology advances.

It is useful to have benchmarks against which computational capacity can be characterized. We will use two: factoring capacity and error probability per quantum logic gate. The former is algorithm dependent, but illustrative, whereas the latter allows us to contrast the physical systems with the error correction threshold estimates for continuous quantum computation.[16] (However, the threshold numbers have been obtained under assumptions that may not be applicable to trapped ions. e.g. an error probability per gate that is independent of the number of qubits.)

## 5. Metastable state qubits

We shall consider a quantum algorithm that requires $L$ qubits (ions), and $n$ laser pulses (we count only the slow, $U$-pulses), each of duration $t$ (a $\pi$-pulse, $\theta = \pi$, for definiteness). Spontaneous emission of just one photon from one of the qubits' $|1\rangle$ states will destroy the quantum coherence required to complete this computation, so we may set an upper limit on the computational time, $nt$, in terms of the spontaneous emission lifetime of this level, $\tau_0$. The specific form of the bound depends on the "average" number of qubits that will occupy the $|1\rangle$ state during the computation: we choose this proportion to be 2/3; giving a bound:

$$nt < 6\tau_0/L \quad . \tag{25}$$

So we see that "more" computation can be performed if the logic gate time, $t$, can be reduced. The duration, $t$, of a $\pi$-pulse is determined by the intensity, $I$, of the laser field:

$t \sim I^{-1/2}$. However, $t$, cannot be made arbitrarily small. In an earlier paper we showed that $t$ cannot be smaller than the period of the CM motion, and shorter periods require stronger axial potentials that push the ions closer together.[15] The shortest possible gate time then corresponds to a minimum ion spacing of one wavelength of the interrogating laser light. In this paper we will consider a different mechanism that gives comparable limits: the breakdown of the two level approximation in intense laser fields, first considered in Reference 30.

In addition to the two states comprising each qubit, there are other ionic levels with higher energies than the |1> state that have rapid electric dipole transitions (lifetime $\tau_{ex}$) to the ground state, and so if some population is transferred to such states during computation their rapid decay will destroy quantum coherence. Although the driving laser frequency is far off-resonance (detuning $\Delta$) from the transition frequency between |0> and a higher lying ("extraneous") level, in intense laser fields there will be some probability, $P$, of occupying this level, given by

$$P \sim \frac{\Omega_{ex}^2}{8\Delta^2} \quad , \tag{26}$$

where $\Omega_{ex}$ is the Rabi frequency for the transition from the ground state, |0>, to the higher lying, extraneous level. Therefore, the probability of decoherence through this two-level breakdown is proportional to the laser intensity, $I$. By requiring that the probability of photon emission from a third level should be less than one during the computation, we obtain the following inequality

$$nt\frac{\Omega_{ex}^2}{8\Delta^2 \tau_{ex}} < 1 \quad . \tag{27}$$

This inequality sets an upper bound on the laser intensity. From the two inequalities (25) and (27) we obtain the bound

$$nL < \eta \left(\frac{20}{\pi}\right)^{1/2} \left(\frac{\lambda_0}{\lambda_{ex}}\right)^{3/2} \tau_{ex}\Delta \quad , \tag{28}$$

between an algorithmic quantity (left-hand side) and a physics parameter (right-hand side), where $\lambda_0$ is the wavelength of the |0> - |1> transition, and $\lambda_{ex}$ is the wavelength of the transition from the extraneous level to the |0> state. Using "typical" values of $\tau_{ex} \sim 10^{-8}$s and $\Delta \sim 10^{15}$ Hz we see that the value of the right-hand-side of this inequality is $\sim \eta.10^7$, translating into enough time to perform a very large number ($10^5$-$10^6$) of logic operations on tens of qubits. (The Lamb-Dicke parameter, $\eta$, for these ions will be $\sim 0.01$-$0.1$.)

The inequality (28) suggests that longer wavelength qubit transitions allow more computation. Indeed, for specific ions we obtain the bounds:

$Hg^+$: $nL < \eta.3 \times 10^7$
$Sr^+$: $nL < \eta.7 \times 10^7$
$Ca^+$: $nL < \eta.1 \times 10^8$
$Ba^+$: $nL < \eta.5 \times 10^8$

suggesting that $Ba^+$ ions may offer greater computational potential than $Hg^+$ or $Ca^+$. However, with $L \sim 60$ qubits the bound (28) in $Ba^+$ corresponds to a computational time $6\tau_0/L \sim 6$ s, whereas technical sources of decoherence such as ion heating and laser phase stability are likely to limit the computation before this limit is reached. Therefore, $Ba^+$ ions are not likely to offer any significant computational advantage over $Ca^+$ at present. We note that when translated into an error probability per gate, the above bounds fail to meet the threshold precision that has been suggested for quantum error correction to allow indefinite quantum computation by one to two orders of magnitude.[13]

## 6. Raman qubits

When qubits are represented by Zeeman or hyperfine sublevels of an ion's ground state, Raman transitions would be used to drive the computational operations, detuned by an amount $\Delta$ below some third level (lifetime $\tau_1$). The Rabi frequency for Raman transitions is proportional to the laser field intensity,

$$\Omega \sim I/\Delta \quad , \tag{29}$$

as is the decoherence process of spontaneous emission from the third level,

$$P \sim I/\Delta^2 \quad . \tag{30}$$

Hence, the probability of a successful computational result is independent of how quickly the computation is performed (at least from the perspective of this decoherence mechanism). Therefore, Raman transitions offer the possibility of completing a computation before technical decoherence mechanisms, such as ion heating, become significant. Using similar arguments as in the last section, we can derive the following inequality for quantum algorithm parameters in terms of the physics parameters for Raman qubits:

$$nL^{1/2} < 8 \, \eta \, \tau_1 \, \Delta \quad . \tag{31}$$

The right-hand side of this inequality has a typical value $\sim \eta.5 \times 10^5$ which is adequate for a large number of gate operations ($\sim 10^6$) on tens of qubits. Also, with the same number of qubits, the error probability per gate is lower for the Raman transitions than with metastable qubits. Therefore, Raman qubits come closer to the error correction thresholds than metastable qubits.[13]

It is also possible to express the computational bounds in terms of typical atomic values of lifetimes, wavelengths etc.[31] However, the bounds obtained this way are considerably more pessimistic than the ones we obtained above because real ions have much longer lived metastable levels than is suggested by the atomic unit of electric quadrupole moment, for instance. Therefore, although indicative of the amount of computation possible, this approach does not provide an absolute upper bound on computational capacity in terms of fundamental constants.[32]

## 7. Quantum factoring with trapped ions

To translate the above physics bounds on algorithmic quantities into limits on the size of integer that could be factored, it is necessary to determine the computational space and time requirements of quantum factoring. Using the values

$$L = 5l+4$$
$$n = 96l^3 + O(l^2)$$

(where $n$ is the number of $U$ pulses) in the decoherence bounds above, we obtain the (algorithm-dependent) factoring limits ($\eta = 0.01$):

$$Hg^+: l < 5 \text{ bits}$$
$$Sr^+: l < 6 \text{ bits}$$
$$Ca^+: l < 6 \text{ bits}$$
$$Ba^+: l < 10 \text{ bits}$$
$$Yb^+: l < 5 \text{ bits}$$

with metastable qubits. (Larger values may be possible with Raman qubits provided a careful optimization of the parameters is made.) These limits correspond roughly to the size of computation at which the probability of success has fallen to $1/e$. Larger integers could be factored but with a lower success probability. Certainly, these projections of the intrinsic factoring capacity of ion trap QCs are insignificant in comparison with the size of integers that are used in cryptography. Nevertheless, the 6-bit factoring limit with $Ca^+$ ions (for instance) represents ~ 20,000 $U$-pulses applied to 34 qubits, taking ~ 0.2 s, representing ample opportunity for studying practical small-scale quantum computation. Also, the above limits do not take into account any possible gains from the use of quantum error correction. We note that the total computational time with metastable qubits is $\sim 6\tau_0/L$, so that it might be possible to reduce the computation time by using an algorithm with additional qubits but less gate operations.

## 8. Summary and Conclusions

In this paper we have discussed the cryptographic significance of quantum computation. Given the enormous disparity between the current state of the art of quantum computation experiments and the requirements for quantum factoring of interesting numbers, it would be easy to dismiss quantum computation as irrelevant for cryptography. However, as we have seen in Section 2, it is the possibility that quantum computation might become possible in 20 years time (say) that must be seriously considered today. Cryptography will therefore be a compelling motivation for quantum computation research.

We have surveyed the prospects for and limitations to quantum computation with trapped ions. It is apparent that with existing technology, adequate time scales and capacity to hold multiple qubits are available to explore quantum computation well beyond the current state-of-the-art: a single logic operation involving two qubits. These intrinsic limits (without quantum error correction) only correspond to the factoring of small integers. However, the numbers of qubits and logic operations involved are huge, and the gate precision with Raman qubits is close to the error correction thresholds for indefinite computation. Ion traps will therefore be a potent method for exploring whether superpositions and entangled states of large numbers of qubits can be created. Investigations of the type studied here identify the relevant physics issues that must be addressed to achieve computational gains. In particular, we note that there has yet to be a demonstration that more than one ion can be sideband cooled to the vibrational ground state. Furthermore, the heating mechanisms for this vibrational mode are poorly understood. [33, 34] Studies of sideband cooling and reheating of multiple ions will therefore be crucial to the development of ion trap QCs. Once entangled states of three or more qubits can be constructed it will also be possible to determine whether multiparticle decoherence mechanisms are consistent with the model that we have used.

We note that ion trap quantum computation offers many advantages to the recently proposed NMR quantum computation model,[28] as summarized in Reference 35. Ion trap qubit coherence is limited by spontaneous emission processes whereas NMR qubit decoherence is thermally dominated ($kT \gg h\nu$). Ion trap quantum information is consequently much more robust. Furthermore, gate times in an ion trap QC could be as short as 1 μs (set by achievable laser intensities, and two-level breakdown), whereas NMR gate times will typically be ~ 0.1-1 s (set by the strength of spin-spin interactions and the need to avoid crosstalk with unintended qubits). Readout in an NMR QC is problematic, with an exponential reduction in magnetization signal with additional qubits, whereas ion trap QC readout is a robust process independent of the number of qubits involved. Moreover, an ion trap QC has the advantage that logic operations can be performed between arbitrary qubits in the register, whereas in NMR only nearest-neighbor operations are possible. Therefore, computation in an NMR QC would use much of the available coherence time in moving qubits around the register until they are adjacent to each other. We estimate that a realistic bound to the computation possible in an NMR QC is about 10 qubits and 100 logic operations. Of these 100 operations many would be used in a typical computation to move separated qubits until they are adjacent. A more detailed comparison of ion trap QCs and NMR will be the subject of a forthcoming paper.[36]

Finally, we should expect that many of the technologies now being pursued for quantum computation will be superceded by even more promising ideas. At present we are probably entering the "vacuum tube" era of this exciting new field.

**Acknowledgements**

It is a pleasure to thank D. F. V. James, S. K. Lamoreaux, M. S. Neergaard and W. Warren for helpful discussions. RJH thanks the ISI Foundation, Torino, Italy for hospitality. This research was funded by the National Security Agency.